\documentclass[apj]{emulateapj}
\usepackage{graphicx}
\usepackage{amssymb}
\usepackage{natbib}

\newcommand{\EQ}{\begin{equation}}
\newcommand{\EN}{\end{equation}}
\newcommand{\nG}{\,{\rm nG}}
\newcommand{\mG}{\,{\rm mG}}

\newcommand{\pc}{\,{\rm pc}}

\slugcomment{draft \today}

\shorttitle{Strong magnetic field generation during first star formation}

\shortauthors{Sur et al.}

\begin{document}

\title{The generation of strong magnetic fields during the 
formation of the first stars}

\author{Sharanya Sur\altaffilmark{1}, D. R. G. Schleicher\altaffilmark{2,3}, 
Robi Banerjee\altaffilmark{1}, Christoph Federrath\altaffilmark{1}, and 
Ralf~ S.~Klessen\altaffilmark{1, 4}}

\altaffiltext{1}{Zentrum f\"ur Astronomie der Universit\"at Heidelberg, \\Institut f\"ur 
Theoretische Astrophysik, Albert-Ueberle-Str.~2, 69120 Heidelberg, Germany}
\altaffiltext{2}{ESO Garching, Karl-Schwarzschild-Str. 2, 85748 Garching bei M\"unchen,
Germany}
\altaffiltext{3}{Leiden Observatory, Leiden University, P.O Box 9513, NL-2300 RA Leiden, 
the Netherlands}
\altaffiltext{4}{Kavli Institute for Particle Astrophysics and Cosmology, Stanford University, 
Menlo Park, CA 94025, USA}

\begin{abstract}
Cosmological hydrodynamical simulations of primordial star formation suggest 
that the gas within the first star-forming halos is turbulent. This has strong implications
on the subsequent evolution, in particular on the generation of magnetic fields. 
Using high-resolution numerical simulations, we show that in the presence of 
turbulence, weak seed magnetic fields are exponentially amplified by the small-scale 
dynamo during the formation of the first stars. 
We conclude that strong magnetic fields are generated during the birth of the first 
stars in the universe, potentially modifying the mass distribution of these stars and 
influencing the subsequent cosmic evolution. We find that the presence of the 
small-scale turbulent dynamo can only be identified in numerical simulations in 
which the turbulent motions in the central core are resolved with at least 32 grid 
cells.
\end{abstract}

\keywords{cosmology: theory --- early universe --- magnetic fields --- turbulence --- 
methods: numerical --- stars: formation}

\section{Introduction}
\label{sec:intro}
Magnetic fields are ubiquitous in the local universe \citep{Beck+96} and
there is growing evidence of their presence also at high redshifts 
\citep{B+08, RQH08, Murphy09}. The seeds for these fields are probably a 
relic from the early universe, possibly arising during inflation or some other 
phase transition \citep[see,][for reviews]{GR01, KS10}. Alternatively, 
they could be generated by astrophysical mechanisms like the Biermann 
battery \citep{B50, Xu+08} or the Weibel instability \citep{SS03, M+04}. 
Regardless of their physical origin, most models predict weak field strengths 
or have very large uncertainties. Thus, magnetic fields are often ignored in 
studies of primordial star formation. 

However, recent cosmological hydrodynamical simulations of first star formation 
\citep{ABN02, OsN07, YOH08, TAOs09} show the presence of turbulence in the 
minihalos where it plays an important role in regulating the transport of angular
 momentum. This opens up the possibility of exponentially amplifying initially 
 weak seed magnetic fields by the action of the small-scale dynamo. 
Field amplification via the small-scale dynamo requires both turbulent gas 
motions and high magnetic Reynolds numbers \citep{BS05}. 

The properties of the small-scale dynamo have been explored both in 
numerical simulations of driven turbulence without self-gravity and in 
analytic models \citep{HBD04, Scheko+04, BS05}, as well as in cosmological 
simulations of magnetic field generation in galaxy clusters \citep{Xu+09}. 
The importance of the small-scale dynamo during galaxy formation was
previously suggested by \citet{BPSS94}, \citet{ABKS09}, and \citet{SO10}. 
It's relevance for primordial star formation was previously assessed by 
\citet{Sch+10}. Possible amplification mechanisms in the protostellar disk have 
been previously identified by \citet{PS89}, \citet{TB04} and \citet{SL06}.

In this Letter, we show, using high resolution numerical simulations that strong 
and dynamically important magnetic fields are generated from initially weak 
seed magnetic fields by the small-scale turbulent dynamo during the collapse 
of primordial halos. The study presented here is a {\em proof-of-concept} 
investigation to show that the small-scale dynamo is operational during the 
collapse of gas clouds, which we apply to the conditions of first star formation. 
In Section~\ref{sec:numerical} we describe our numerical method and the 
initial conditions. The results obtained from our numerical simulations are 
presented in Section~\ref{sec:results}. Finally, we summarize and discuss 
our findings in Section~\ref{sec:conc}.

\section{Numerical Method and Initial conditions}
\label{sec:numerical}
We focus on the gravitational collapse and magnetic field amplification of the 
baryon-dominated inner parts of a contracting primordial halo, using a simplified 
initial setup (rather than a full cosmological simulation) where we neglect the 
detailed thermodynamics and non-ideal MHD effects.
The numerical simulations presented here were performed with the publicly 
available adaptive-mesh refinement (AMR) code, FLASH2.5 \citep{Fryxell00}. 
We solve the equations of ideal MHD including self-gravity with a refinement 
criterion guaranteeing that the Jeans length, 
\begin{equation}
\label{eq:jlength}
\lambda_{\rm J} = \left(\frac{\pi\,c_{\rm s}^{2}}{G\,\rho}\right)^{1/2}\,,
\end{equation}
with the sound speed $c_{\rm s}$ and the gravitational constant $G$, is 
always resolved with a user-defined number of cells. We use an MHD 
Riemann solver developed for FLASH that preserves positive states and 
proved to be highly efficient and accurate for modeling astrophysical problems 
involving turbulence and strong shocks \citep{BKW07, W09, BKW10, WFK10}.  

The initial conditions for our numerical simulations were motivated from 
larger-scale cosmological models of \citet{ABN02}, \citet{BCL02} and 
\citet{YOH08}. We initialize a super-critical Bonnor-Ebert (BE) sphere with a 
core density of $\rho_{\rm BE} \simeq 4.68\times 10^{-20}\, {\rm g\, cm}^{-3}$ 
($n_{\rm BE} = 10^4\,{\rm cm}^{-3}$) at a temperature of $T = 300\, {\rm K}$.
We also include a small amount of rotation of about $4\%$ of the gravitational 
energy. The radius of the BE-sphere is $1.5 \,\pc$ which corresponds to a 
dimensionless radius of $\xi = 8.28$. Note that the critical value is $6.451$ 
\citep{Eb55, Bonnor56}. The total computational domain is $(3.9\, \pc)^3$ in 
size. The initial conditions are furthermore characterized by the presence of 
a random initial velocity field with transonic velocity dispersion of amplitude 
$1.1\,{\rm km}\,{\rm s}^{-1}$ (equal to the initial sound speed) and a weak 
random magnetic field with $B_{\rm rms} \sim 1\,{\nG}$. 
This initial field strength corresponds to $\beta\approx 10^{10}$, where $\beta$ 
is the ratio of the thermal pressure to magnetic pressure. Both, the turbulent 
energy and magnetic field spectra were initialized with the same power law 
dependence in wave number space, $\propto k^{-2}$, with most power on scales 
$\sim\,0.8\,{\rm pc}$, which roughly corresponds to the initial Jeans length of the core. 
Consistent with previous works of \citet{Omukai05} and \citet{GS09}, which follow 
the thermodynamics during the collapse, we adopt an effective equation of state 
with $\Gamma=d\log T / d\log\rho + 1 = 1.1$ for number densities in the range 
$n = 10^{5} - 10^{10}\, {\rm cm}^{-3}$. 
These initial conditions reflect the physical parameters of the baryon-dominated 
regime in the centers of the first minihalos~\citep{ABN02}. Hence we can safely 
neglect the influence of dark matter at this point\footnote{Dark matter would 
only have an indirect influence on the magnetic field dynamics through its 
gravitational interaction with the plasma. At low densities, it may accelerate the 
collapse and introduce additional turbulence in the gas, which could potentially 
enhance the dynamo.}. The applicability of the ideal MHD approximation for the 
description of the dynamics has been studied by \citet{MS04} using one-zone 
models that follow the chemistry, and in particular the abundances of ionized 
species during the protostellar collapse phase. Their study suggests that the 
ionization degree is sufficiently high to ensure a strong coupling between ions 
and neutrals to maintain flux-freezing.

We note that the efficiency of the dynamo process depends on the Reynolds 
number and is thus related to how well the turbulent motions are resolved 
\citep{HBD04, Balsara+04}. Higher resolution yields larger field amplification. 
To demonstrate this effect, we perform five numerical simulations where we 
resolve $\lambda_{\rm J}$ by 8, 16, 32, 64 and 128 cells.  

\section{Results}
\label{sec:results}
\begin{figure}
\begin{center}
\begin{tabular}{c}
\includegraphics[width=0.95\columnwidth]{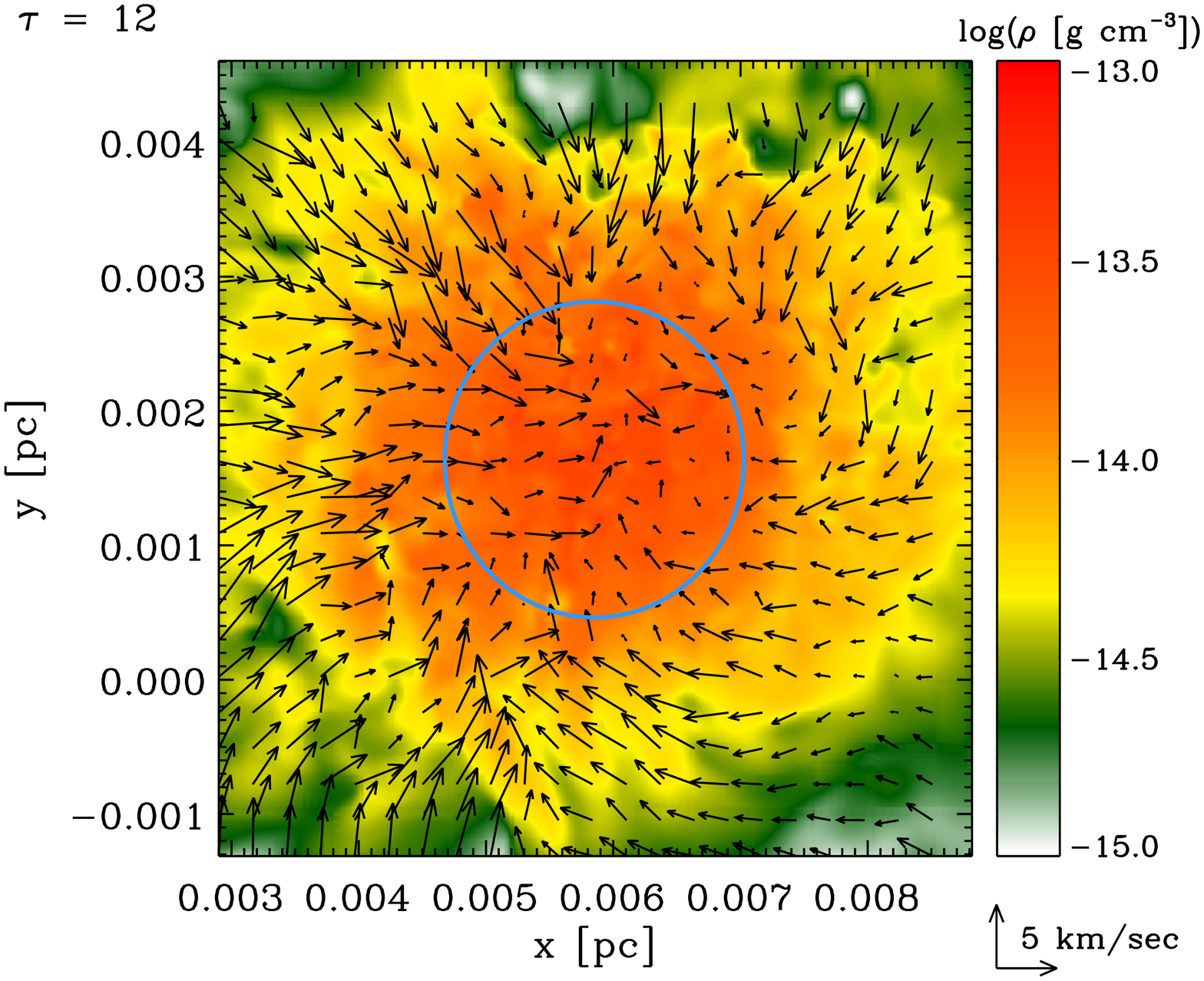} \\
\includegraphics[width=0.95\columnwidth]{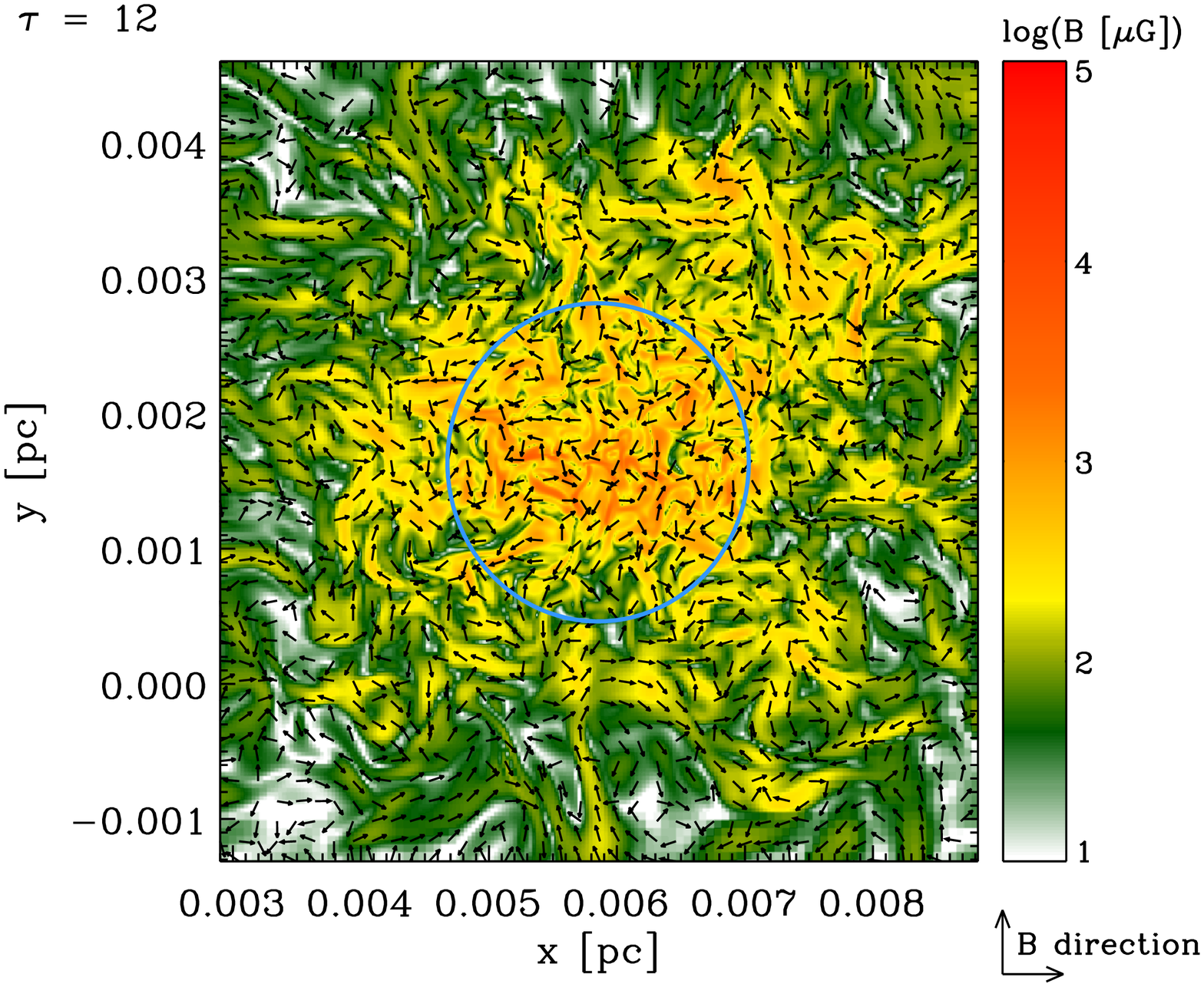}
\end{tabular}
\end{center}
\caption{Two dimensional slices through the center of the collapsing core 
at the time when the initial field strength has increased by a factor of $\sim\!10^6$, 
showing the central region of our highest-resolution simulation ($\lambda_{\rm J}$ 
resolved by 128 cells). The circle indicates the control volume $V_{\rm J}$ 
centered on the position of the current density peak. 
The top image shows the density and the velocity component in the $xy$-plane, 
indicating radial infall in the outer regions and turbulent motions in the inner core. 
The bottom image depicts the total magnetic field amplitude and the local 
magnetic field direction. 
\label{slices}}
\end{figure}

We find that the dynamical evolution of the system is characterized by two distinct 
phases. First, as the initial turbulent velocity field decays the system exhibits weak 
oscillatory behavior and contracts only slowly. Soon, however, the runaway collapse 
sets in. Fig.~\ref{slices} shows a snapshot of the central region of the collapsing core 
in our highest-resolution simulation at a time when the central density has increased 
by a factor of $\sim 10^{6}$. The magnetic field strength has grown by a factor of 
$10^{6}$, reaching a peak value of about $1\, \mG$ at $\tau\sim 12$. The top image 
shows the density and the velocity structure, and the bottom image shows the 
magnetic field strength and the local magnetic field directions. 

\begin{figure}
\centerline{\includegraphics[width=0.99\columnwidth]{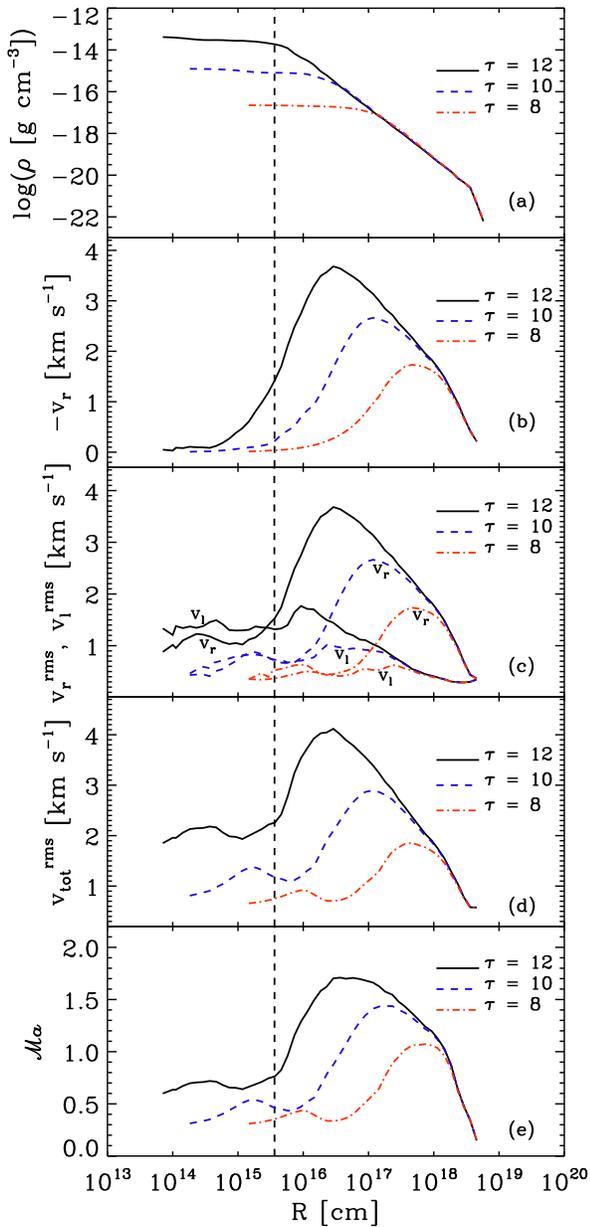}}
\caption{Time evolution of the radial density profile (panel a),
  the radial profile of the mean value of the infall velocity, $v_{\rm r}$ (panel b), 
  the radial profile of the rms value of $v_{\rm r}$  
  together with the rms fluctuations of the lateral velocity $v_{\rm l}$, 
  i.e., the component perpendicular to $v_{\rm r}$ (panel c), the radial profile 
  of the rms value of the total velocity (panel d) and the Mach number $\mathcal{M}$$a$
  (panel e). The vertical line indicates the Jeans radius at $\tau = 12$. 
  The data are taken from our highest-resolution run.
  \label{radial}}
\end{figure}

To understand the behavior of the system more quantitatively, we need to 
follow its dynamical contraction. First, we note that the physical time scale 
becomes progressively shorter during collapse. We therefore define a 
dimensionless time coordinate $\tau$, 
\EQ
\label{tau}
\tau = \int dt / t_{\rm ff} (t)\,,
\EN
which is normalized in terms of the local free-fall time, 
\EQ
\label{tff}
t_{\rm ff}(t)=\left(\frac{3\pi}{32\,G\rho_{\rm m}(t)}\right)^{1/2},
\EN 
where $\rho_{\rm m}(t)$ is the mean density of the contracting central region. 
We also define a control volume for gravitational collapse based on the Jeans 
volume, $V_{\rm J} = 4\pi  (\lambda_{\rm J}/2)^3/3$. We obtain all dynamical 
quantities of interest as averages within the contracting Jeans volume, 
which is centered on the position of the maximum density. This approach 
ensures that we always average over the relevant volume for collapse and 
field amplification.

The density profile always maintains a flat inner region within the size of the 
local Jeans length. This is illustrated in Fig.~\ref{radial}a at three different times, 
where a flat core of the size of $\lambda_{\rm J}$ is sustained throughout the 
collapse. The radial infall motions dominate the total velocity in the envelope
as shown in Fig.~\ref{radial}b. The turbulence is maintained on scales 
{\it below} $\lambda_{\rm J}$ and dominates the dynamics inside the core 
region as can be seen from Figs.~\ref{radial}c and \ref{radial}d, where we show 
the radial profiles of the magnitude of the fluctuation (rms) value of the infall 
velocity together with the rms of the lateral component of the velocity and the 
rms value of the total velocity, respectively. We show in Fig.~2e, the radial 
profile of the Mach number which illustrates that the infalling velocities are 
supersonic, while inside the core, the velocities are subsonic.

\begin{figure}
\centerline{\includegraphics[width=0.99\columnwidth]{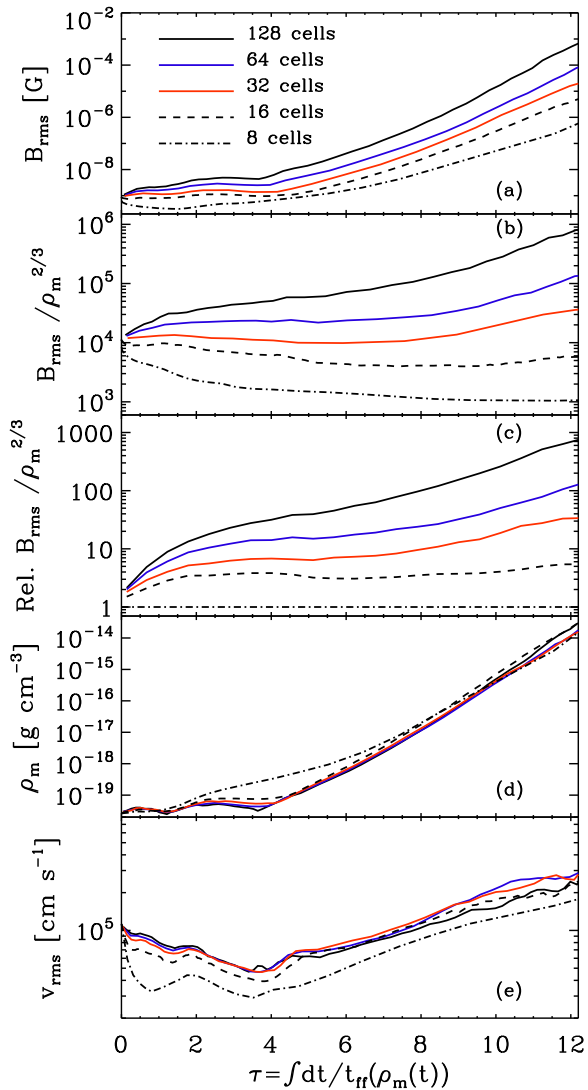}}
\caption{Evolution of the dynamical quantities in the central Jeans volume 
as a function of $\tau$, defined in equation~(\ref{tau}) for five runs with different 
number of cells to resolve the local Jeans length. Panel (a) shows the rms 
magnetic field strength $B_{\rm rms}$, amplified to $1\,$mG from an initial 
field strength of $1\,$nG, (b) the evolution of $B_{\rm rms}/\rho_{\rm m}^{2/3}$, 
showing the turbulent dynamo amplification by dividing out the maximum 
possible amplification due to perfect flux freezing, (c) the relative amplification 
in $B_{\rm rms}/\rho_{\rm m}^{2/3}$ compared to the 8 cell run, 
(d) the evolution of the mean density $\rho_{\rm m}$ and (e) the rms velocity 
$v_{\rm rms}$. The runaway collapse commences at about $\tau\sim 4$. 
\label{evol}}
\end{figure}

\begin{figure}
\centerline{\includegraphics[width=0.95\columnwidth]{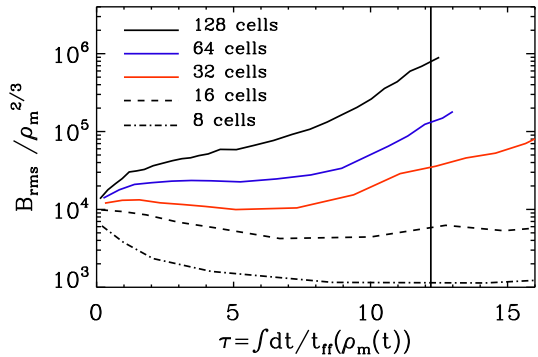}}
\caption{The figure illustrates the minimum resolution criterion required to capture 
the growth of the magnetic field due to small-scale dynamo action. The dynamo
begins to be observed for simulations where $\lambda_{\rm J}$ is resolved by
a minimum of 32 cells. Simulations performed with the Jeans length resolved 
by either 8 cells or 16 cells are decaying in nature with weak fluctuations. 
The vertical line indicates the values of $B_{\rm rms}/\rho_{\rm m}^{2/3}$ obtained 
in the different resolution runs at $\tau=12.2$. 
\label{lowres}}
\end{figure}

Gravitational compression during the collapse of a primordial gas cloud can at most 
lead to an amplification of the magnetic field strength by a factor of $\sim \rho^{2/3}$ 
in the limit of perfect flux freezing (i.e., ideal MHD). A stronger increase implies the 
presence of an additional amplification mechanism. Starting from an initial field 
strength of $\sim 1{\rm nG}$, our simulations show a total magnetic field amplification 
by six orders of magnitude, leading to a field strength of about $\sim 1\,$mG for the 
case where we resolve the local Jeans length by 128 cells. 
This is illustrated in Fig.~\ref{evol}a. Fig.~\ref{evol}b shows that in our highest resolution 
simulation, the obtained field amplification is indeed stronger than what is expected from 
pure flux freezing, which demonstrates that the small-scale turbulent dynamo provides 
significant additional field amplification over compression. 
Fig.~\ref{evol}c shows a plot of the relative amplification in simulations with 16, 32, 
64 and 128 cells compared to the simulation where $\lambda_{\rm J}$ is resolved by 8 
cells, i.e., we divided each curve in Fig.~\ref{evol}b by the curve of the 8 cell run.
The time evolution of the mean density $\rho_{\rm m}$ within the central Jeans 
volume is depicted in Fig.~\ref{evol}d, while Fig.~\ref{evol}e shows the plot of the 
rms velocity. The presence of turbulence delays the collapse until $\tau\sim 4$. 
During this time, the mean density shows oscillations while the rms velocity 
decreases as the turbulence decays. 
Sufficient numerical resolution is a crucial issue when studying the small-scale turbulent 
dynamo. This is evident from Fig.~\ref{lowres}, where the simulations with 
$\lambda_{\rm J}$ resolved by 8, 16 and 32 cells are evolved further till $\tau=16$. 
The 8 and 16 cell runs do not show any increase in $B_{\rm rms}/\rho_{\rm m}^{2/3}$ and 
only exhibit weak fluctuations, while the 32 cell run is the first to show an increase. 
This signifies that a minimum of 32 cells per Jeans 
length is required to obtain the exponential amplification of the magnetic field by the 
small-scale dynamo. This is also 
consistent with the turbulence simulations by \citet{Federrath+10}, who concluded that 
at least 30 grid cells are required to resolve turbulent vortices.

As mentioned in Section~2, the growth rate of the magnetic field depends on the Reynolds 
number of the system and is thus related to the numerical resolution of the simulation. 
With increasing Reynolds number and thus with higher numerical resolution, the growth 
rate of the dynamo-generated magnetic field increases \citep{HBD04, Balsara+04}. 
In agreement with this, our resolution study shows a divergent behavior in the growth rate 
of the magnetic field (Fig.~\ref{evol}b and \ref{evol}c) rather than a convergent one. 
A convergence in the growth rate can only be obtained if the physical length scales that 
determine the Reynolds number are sufficiently resolved. Our simulations also show no 
signs of saturation and thus we simply stop the calculation when the numerical cost 
becomes prohibitively high. In reality, we expect the field amplification to continue until 
back-reactions either via the Lorentz force \citep{S99,Scheko+04} or via non-ideal MHD 
effects such as ambipolar diffusion \citep{PG08} become important. Calculations of 
MHD turbulence without self-gravity indicate maximum field strengths within $5$\% of the 
equipartition value \citep{S99}. We note that the physical dissipation scales for ambipolar 
diffusion and Ohmic dissipation are much smaller than the Jeans length and thus the 
growth rates obtained in our simulations are lower limits to the physical growth rates. 

\section{Discussion and Conclusions}
\label{sec:conc}
Studies of primordial star formation show that the gas within the first star forming 
halos is turbulent \citep{ABN02, OsN07, YOH08, TAOs09}. In this Letter, we showed 
that this turbulence is maintained throughout the collapse and drives a small-scale 
dynamo, which exponentially amplifies weak magnetic seed fields. We performed five 
numerical simulations of collapsing Bonnor-Ebert spheres where the local Jeans length 
was resolved by 8, 16, 32, 64 and 128 cells. Our numerical simulations show that, 
starting from a weak seed field of $\sim 1\, \nG$, strong and dynamically important fields 
can eventually be generated in the central collapsing core during the formation of the 
first stars. The small-scale dynamo only works in simulations in which the turbulent 
motions in the central core are sufficiently resolved. We find that a minimum resolution 
of 32 cells per Jeans length is necessary for resolving the small-scale dynamo in 
our simulations. 

The generation of strong and dynamically important magnetic fields has interesting 
consequences for our understanding of how the first stars form and how they 
influence subsequent cosmic evolution. We know from modeling galactic 
star-forming clouds that the presence of magnetic fields can reduce the level of 
fragmentation, and by doing so strongly influences the stellar mass spectrum 
\citep{HT08}. Furthermore, the dynamo-generated strong magnetic fields can drive 
jets and outflows from the accretion disks via the magneto-centrifugal mechanism 
\citep{RBDS03}. Such outflows remove a significant fraction of the mass and 
angular momentum which influences the stellar mass spectrum. The implications 
of magnetic fields in self-gravitating disks have been explored by \citet{Fromang+04}, 
who find that the interaction of turbulence excited by the magneto-rotational instability 
with the self-gravitational instability excites additional modes, broadens the spiral 
arms and effectively reduces the accretion rate due to non-linear interaction. There 
are first attempts to study magnetic fields in the context of first star formation 
\citep{Machida+06}, but more sophisticated initial conditions and more appropriate 
magnetic field geometries need to be considered.

Once the first stars have formed, they are likely to produce a copious amount of ionizing 
photons, which drive huge H$\,\textsc{ii}$ regions, bubbles of ionized gas, expanding 
into the low-density gas between the halos. The effects of radiation have been shown 
in contemporary star formation \citep{KMc05, KKMc07, Peters+10a, Peters+10b}, and are 
likely to be important for the first stars as well \citep{WA08, GJKB09}.
The expansion of the H$\,\textsc{ii}$ region could be substantially different if magnetic 
outflows drive a cavity into the surrounding gas. The magnetic field may further affect 
fluid instabilities near the ionization front.

The mechanism of exponentially amplifying weak seed magnetic fields via the 
small-scale dynamo is likely to work not only during the formation of the first stars, 
but in all types of gravitationally bound, turbulent objects. Highly magnetized gas is 
thus expected already in the first galaxies.   

\acknowledgments{S.~Sur thanks the German Science Foundation (DFG) for 
financial support via the priority program 1177 'Witnesses of Cosmic History: 
Formation and evolution of black holes, galaxies and their environment' 
(grant KL 1358/10). D.~R.~G Schleicher is supported by the European 
Community's Seventh Framework Programme (FP7/2007-2013) under 
grant agreement No 229517. R. Banerjee is funded by the Emmy-Noether 
grant (DFG) BA 3607/1. C.~Federrath and R.~S.~Klessen are supported 
by the Landesstiftung Baden-W{\"u}rttemberg via their program International 
Collaboration II under grant P-LS-SPII/18. R.~S.~K thanks the KIPAC 
at Stanford University and the Department of Astronomy and Astrophysics
at the University of California at Santa Cruz for their warm hospitality during 
a sabbatical stay in 2010. The KIPAC is sponsored in part by the U.~S. 
Department of Energy contract no. DE - AC - 02 - 76SF00515.
We acknowledge computing time at the Leibniz-Rechenzentrum in Garching 
(Germany) and partial support from a Frontier grant of Heidelberg University 
funded by the German Excellence Initiative. The FLASH code is developed in 
part by the DOE-supported Alliances Center for Astrophysical Thermonuclear 
Flashes (ASC) at the University of Chicago. 
}


\begin{thebibliography}{50}
\expandafter\ifx\csname natexlab\endcsname\relax\def\natexlab#1{#1}\fi

\bibitem[{{Abel} {et~al.}(2002){Abel}, {Bryan}, \& {Norman}}]{ABN02}
{Abel}, T., {Bryan}, G.~L., \& {Norman}, M.~L. 2002, Science, 295, 93

\bibitem[{{Arshakian} {et~al.}(2009){Arshakian}, {Beck}, {Krause}, \&
  {Sokoloff}}]{ABKS09}
{Arshakian}, T.~G., {Beck}, R., {Krause}, M., \& {Sokoloff}, D. 2009, \aap,
  494, 21

\bibitem[{{Balsara} {et~al.}(2004){Balsara}, {Kim}, {Mac Low}, \&
  {Mathews}}]{Balsara+04}
{Balsara}, D.~S., {Kim}, J., {Mac Low}, M., \& {Mathews}, G.~J. 2004, \apj,
  617, 339

\bibitem[{{Beck} {et~al.}(1996){Beck}, {Brandenburg}, {Moss}, {Shukurov}, \&
  {Sokoloff}}]{Beck+96}
{Beck}, R., {Brandenburg}, A., {Moss}, D., {Shukurov}, A., \& {Sokoloff}, D.
  1996, \araa, 34, 155

\bibitem[{{Beck} {et~al.}(1994){Beck}, {Poezd}, {Shukurov}, \&
  {Sokoloff}}]{BPSS94}
{Beck}, R., {Poezd}, A.~D., {Shukurov}, A., \& {Sokoloff}, D.~D. 1994, \aap,
  289, 94

\bibitem[{{Bernet} {et~al.}(2008){Bernet}, {Miniati}, {Lilly}, {Kronberg}, \&
  {Dessauges-Zavadsky}}]{B+08}
{Bernet}, M.~L., {Miniati}, F., {Lilly}, S.~J., {Kronberg}, P.~P., \&
  {Dessauges-Zavadsky}, M. 2008, \nat, 454, 302

\bibitem[{{Biermann}(1950)}]{B50}
{Biermann}, L. 1950, Zeitschrift Naturforschung Teil A, 5, 65

\bibitem[{{Bonnor}(1956)}]{Bonnor56}
{Bonnor}, W.~B. 1956, \mnras, 116, 351

\bibitem[{{Bouchut} {et~al.}(2007){Bouchut}, {Klingenberg}, \&
  {Waagan}}]{BKW07}
{Bouchut}, F., {Klingenberg}, C., \& {Waagan}, K. 2007, Numerische Mathematik,
  108, 7

\bibitem[{{Bouchut} {et~al.}(2010){Bouchut}, {Klingenberg}, \&
  {Waagan}}]{BKW10}
---. 2010, Numerische Mathematik, 115, 647

\bibitem[{{Brandenburg} \& {Subramanian}(2005)}]{BS05}
{Brandenburg}, A., \& {Subramanian}, K. 2005, \physrep, 417, 1

\bibitem[{{Bromm} {et~al.}(2002){Bromm}, {Coppi}, \& {Larson}}]{BCL02}
{Bromm}, V., {Coppi}, P.~S., \& {Larson}, R.~B. 2002, \apj, 564, 23

\bibitem[{{de Souza} \& {Opher}(2010)}]{SO10}
{de Souza}, R.~S., \& {Opher}, R. 2010, \prd, 81, 067301

\bibitem[{{Ebert}(1955)}]{Eb55}
{Ebert}, R. 1955, Zeitschrift fur Astrophysik, 37, 217

\bibitem[{{Federrath} {et~al.}(2010){Federrath}, {Roman-Duval}, {Klessen},
  {Schmidt}, \& {Mac Low}}]{Federrath+10}
{Federrath}, C., {Roman-Duval}, J., {Klessen}, R.~S., {Schmidt}, W., \& {Mac
  Low}, M. 2010, \aap, 512, A81+

\bibitem[{{Fromang} {et~al.}(2004){Fromang}, {Balbus}, {Terquem}, \& {De
  Villiers}}]{Fromang+04}
{Fromang}, S., {Balbus}, S.~A., {Terquem}, C., \& {De Villiers}, J. 2004, \apj,
  616, 364

\bibitem[{{Fryxell} {et~al.}(2000){Fryxell}, {Olson}, {Ricker}, {Timmes},
  {Zingale}, {Lamb}, {MacNeice}, {Rosner}, {Truran}, \& {Tufo}}]{Fryxell00}
{Fryxell}, B., {Olson}, K., {Ricker}, P., {Timmes}, F.~X., {Zingale}, M.,
  {Lamb}, D.~Q., {MacNeice}, P., {Rosner}, R., {Truran}, J.~W., \& {Tufo}, H.
  2000, \apjs, 131, 273

\bibitem[{{Glover} \& {Savin}(2009)}]{GS09}
{Glover}, S.~C.~O., \& {Savin}, D.~W. 2009, \mnras, 393, 911

\bibitem[{{Grasso} \& {Rubinstein}(2001)}]{GR01}
{Grasso}, D., \& {Rubinstein}, H.~R. 2001, \physrep, 348, 163

\bibitem[{{Greif} {et~al.}(2009){Greif}, {Johnson}, {Klessen}, \&
  {Bromm}}]{GJKB09}
{Greif}, T.~H., {Johnson}, J.~L., {Klessen}, R.~S., \& {Bromm}, V. 2009,
  \mnras, 399, 639

\bibitem[{{Haugen} {et~al.}(2004){Haugen}, {Brandenburg}, \& {Dobler}}]{HBD04}
{Haugen}, N.~E., {Brandenburg}, A., \& {Dobler}, W. 2004, \pre, 70, 016308

\bibitem[{{Hennebelle} \& {Teyssier}(2008)}]{HT08}
{Hennebelle}, P., \& {Teyssier}, R. 2008, \aap, 477, 25

\bibitem[{{Krumholz} {et~al.}(2007){Krumholz}, {Klein}, \& {McKee}}]{KKMc07}
{Krumholz}, M.~R., {Klein}, R.~I., \& {McKee}, C.~F. 2007, \apj, 656, 959

\bibitem[{{Krumholz} {et~al.}(2005){Krumholz}, {McKee}, \& {Klein}}]{KMc05}
{Krumholz}, M.~R., {McKee}, C.~F., \& {Klein}, R.~I. 2005, \apjl, 618, L33

\bibitem[{{Machida} {et~al.}(2006){Machida}, {Omukai}, {Matsumoto}, \&
  {Inutsuka}}]{Machida+06}
{Machida}, M.~N., {Omukai}, K., {Matsumoto}, T., \& {Inutsuka}, S. 2006, \apj,
  647, L1

\bibitem[{{Maki} \& {Susa}(2004)}]{MS04}
{Maki}, H., \& {Susa}, H. 2004, \apj, 609, 467

\bibitem[{{Medvedev} {et~al.}(2004){Medvedev}, {Silva}, {Fiore}, {Fonseca}, \&
  {Mori}}]{M+04}
{Medvedev}, M.~V., {Silva}, L.~O., {Fiore}, M., {Fonseca}, R.~A., \& {Mori},
  W.~B. 2004, Journal of Korean Astronomical Society, 37, 533

\bibitem[{{Murphy}(2009)}]{Murphy09}
{Murphy}, E.~J. 2009, \apj, 706, 482

\bibitem[{{Omukai} {et~al.}(2005){Omukai}, {Tsuribe}, {Schneider}, \&
  {Ferrara}}]{Omukai05}
{Omukai}, K., {Tsuribe}, T., {Schneider}, R., \& {Ferrara}, A. 2005, \apj, 626,
  627

\bibitem[{{O'Shea} \& {Norman}(2007)}]{OsN07}
{O'Shea}, B.~W., \& {Norman}, M.~L. 2007, \apj, 654, 66

\bibitem[{{Peters} {et~al.}(2010{\natexlab{a}}){Peters}, {Banerjee}, {Klessen},
  {Mac Low}, {Galv{\'a}n-Madrid}, \& {Keto}}]{Peters+10a}
{Peters}, T., {Banerjee}, R., {Klessen}, R.~S., {Mac Low}, M.,
  {Galv{\'a}n-Madrid}, R., \& {Keto}, E.~R. 2010{\natexlab{a}}, \apj, 711, 1017

\bibitem[{{Peters} {et~al.}(2010{\natexlab{b}}){Peters}, {Mac Low}, {Banerjee},
  {Klessen}, \& {Dullemond}}]{Peters+10b}
{Peters}, T., {Mac Low}, M., {Banerjee}, R., {Klessen}, R.~S., \& {Dullemond},
  C.~P. 2010{\natexlab{b}}, \apj, 719, 831

\bibitem[{{Pinto} \& {Galli}(2008)}]{PG08}
{Pinto}, C., \& {Galli}, D. 2008, \aap, 484, 17

\bibitem[{{Pudritz} \& {Silk}(1989)}]{PS89}
{Pudritz}, R.~E., \& {Silk}, J. 1989, \apj, 342, 650

\bibitem[{{Robishaw} {et~al.}(2008){Robishaw}, {Quataert}, \& {Heiles}}]{RQH08}
{Robishaw}, T., {Quataert}, E., \& {Heiles}, C. 2008, \apj, 680, 981

\bibitem[{{Schekochihin} {et~al.}(2004){Schekochihin}, {Cowley}, {Taylor},
  {Maron}, \& {McWilliams}}]{Scheko+04}
{Schekochihin}, A.~A., {Cowley}, S.~C., {Taylor}, S.~F., {Maron}, J.~L., \&
  {McWilliams}, J.~C. 2004, \apj, 612, 276

\bibitem[{{Schleicher} {et~al.}(2010){Schleicher}, {Banerjee}, {Sur},
  {Arshakian}, {Klessen}, {Beck}, \& {Spaans}}]{Sch+10}
{Schleicher}, D.~R.~G., {Banerjee}, R., {Sur}, S., {Arshakian}, T.~G.,
  {Klessen}, R.~S., {Beck}, R., \& {Spaans}, M. 2010, \mnras, submitted
  (arXiv:1003.1135)

\bibitem[{{Schlickeiser} \& {Shukla}(2003)}]{SS03}
{Schlickeiser}, R., \& {Shukla}, P.~K. 2003, \apj, 599, L57

\bibitem[{{Silk} \& {Langer}(2006)}]{SL06}
{Silk}, J., \& {Langer}, M. 2006, \mnras, 371, 444

\bibitem[{{Subramanian}(1999)}]{S99}
{Subramanian}, K. 1999, \prl, 83, 2957

\bibitem[{{Subramanian}(2010)}]{KS10}
---. 2010, Astronomische Nachrichten, 331, 110

\bibitem[{{Tan} \& {Blackman}(2004)}]{TB04}
{Tan}, J.~C., \& {Blackman}, E.~G. 2004, \apj, 603, 401

\bibitem[{{Turk} {et~al.}(2009){Turk}, {Abel}, \& {O'Shea}}]{TAOs09}
{Turk}, M.~J., {Abel}, T., \& {O'Shea}, B. 2009, Science, 325, 601

\bibitem[{{von Rekowski} {et~al.}(2003){von Rekowski}, {Brandenburg}, {Dobler},
  \& {Shukurov}}]{RBDS03}
{von Rekowski}, B., {Brandenburg}, A., {Dobler}, W., \& {Shukurov}, A. 2003,
  \aap, 398, 825

\bibitem[{{Waagan}(2009)}]{W09}
{Waagan}, K. 2009, Journal of Computational Physics, 228, 8609

\bibitem[{{Waagan} {et~al.}(2010){Waagan}, {Federrath}, \&
  {Klingenberg}}]{WFK10}
{Waagan}, K., {Federrath}, C., \& {Klingenberg}, C. 2010, Journal of
  Computational Physics, submitted

\bibitem[{{Wise} \& {Abel}(2008)}]{WA08}
{Wise}, J.~H., \& {Abel}, T. 2008, \apj, 685, 40

\bibitem[{{Xu} {et~al.}(2009){Xu}, {Li}, {Collins}, {Li}, \& {Norman}}]{Xu+09}
{Xu}, H., {Li}, H., {Collins}, D.~C., {Li}, S., \& {Norman}, M.~L. 2009, \apj,
  698, L14

\bibitem[{{Xu} {et~al.}(2008){Xu}, {O'Shea}, {Collins}, {Norman}, {Li}, \&
  {Li}}]{Xu+08}
{Xu}, H., {O'Shea}, B.~W., {Collins}, D.~C., {Norman}, M.~L., {Li}, H., \&
  {Li}, S. 2008, \apj, 688, L57

\bibitem[{{Yoshida} {et~al.}(2008){Yoshida}, {Omukai}, \& {Hernquist}}]{YOH08}
{Yoshida}, N., {Omukai}, K., \& {Hernquist}, L. 2008, Science, 321, 669

\end{thebibliography}

\end{document}